# Prototyping a Virtual Agent for Pre-school English Teaching

Eduardo Benitez Sandoval*
University of New South Wales
Fac. Arts Design and Architecture Sydney, Australia

Diego Vazquez Rojas
Clarissa A. Parada Cereceres
Alvaro Anzueto Rios[†]
Instituto Politecnico Nacional UPIITA, Mexico

Amit Barde
Mark Billinghurst[‡]
University of Auckland, ECL, New Zealand
UniSA, Australia

**ABSTRACT**

This paper describes a case study and the insights gained from prototyping an Intelligent Virtual Agent (IVA) for English vocabulary building for Spanish-speaking preschool children. After an initial exploration to evaluate the feasibility of developing an IVA, we followed a Human-Centered Design (HCD) approach to create a prototype. We report on the multidisciplinary process used that incorporated two well-known educative concepts: gamification and story-telling as the main components for engagement. Our results suggest that a multidisciplinary approach to developing an educational IVA is effective. We report on the relevant aspects of the ideation and design processes that informed the vision and mission of the project.

## 1 INTRODUCTION

In a global context, expanding language skills to speak a global language (English) could allow better opportunities in education, business, research and many other industries. Interactive technologies such as social robots and Intelligent Virtual Agents (IVA) could significantly support teachers engaged in teaching English as a Second Language. Social agents could enhance current language teaching, providing tailored instruction, motivation, feedback, and tracking to students. The importance of these functions in the learning process has motivated the development of the IVA described in this paper. We established a Human-Centred Design (HCD) approach to avoid a techno-deterministic approach [1].

Our contributions can be described in two parts; a) Reporting the tensions and struggles of a multidisciplinary design process among engineers, educators, and designers has demonstrated the process to be effective, b) Proving the first example of an IVA developed for preschool foreign language learning in the Spanish language context. Reporting the qualitative design process is a novel contribution with enormous potential for remote teaching of relevant skills for preschool children in Spanish speaking regions.

The ability to learn a foreign language differs from person to person. The cognitive process of learning a new language is still not well-known. However, Hinton et al. [2] suggest that motivation plays an intrinsic role in the effects that teaching strategies have on learning. Thereby, the higher the student's motivation, the better and more effective the learning experience. In our research we are interested in how IVA could increase student motivation for language learning. Through our work, researchers interested in multidisciplinary IVAs design could become more aware of the complexities involved in the process, the motivations, and the discussions required to benefit the final user. Using heuristic evaluation and small pilot studies to guide the design of an IVA for language learning could lead to better quantitative evaluations. This approach is not reported often and can be used as a guideline for IVA designs with higher chances of acceptability in the future.

## 2 IMPLEMENTATION: 1ST AND 2ND ITERATIONS.

Preschool children communicate naturally using speech, and use hearing and vision to understand the world around them. However, the learning of a new language in a mono-cultural environment is challenging for them. The goal of an educative multi-modal virtual agent is to achieve natural interaction taking children's capabilities and classroom limitations into account. Under these considerations, we developed an initial prototype that supports novel multi-modal interactions in the language education domain for Spanish speaking preschool children. The focus was on speech recognition and object identification.

The initial prototype consisted of an object recognition module and a speech recognition module. We then added an IVA and a synthetic voice module. In the first iteration the focus was on practising specific words. Initially, ten English words were selected from the typical objects in the everyday life of preschool children. For instance, fruits, toys, and kitchen utensils. However, when tested with children, the evidence collected on video suggested that they were not engaged in the learning process at all. The did not appear to be having fun or learning the new English words either.

Between the first and second iterations, we added a rudimentary IVA capable of basic interaction with a children developed in the Unity game engine. See Figure 1. After the first round of interactions using the IVA with an infant, it was evident that changes in design and delivery methods were required to the initial prototype. For the second iteration, three brainstorming sessions led us to explore the use of gamification and storytelling to engage the children. This demonstrated to us that there were issues beyond technical problems that needed addressing to ensure that the IVA could be useful. This is when an HCD approach was established in the development of the project and the second iteration initiated. After several ideation sessions, we delivered a prototype which incorporates the concepts of gamification and story-telling. This integrated a virtual character, speech recognition module and object recognition module as shown in Fig.2. The children interacted with the prototype by playing "I spy"in a domestic context (helping the IVA to get things from the market for her mother), and collecting points when the pronunciation of the English word was correct (using three attempts). The gamification process improved the engagement of users by including game play elements in the interaction [3]. In this prototype, storytelling was used to enhance user participation and provide the appropriate visual and auditory context to enable faster learning.

## 3 STUDY DESIGN AND VALIDATION

We shaped a design process based on two iterative steps. Each step contains a validation stage to refine the prototype in terms of usability. The first phase of the validation process was comprised of qualitative testing with two preschool children. We recruited two children, a boy and a girl, of preschool age (4-5 years old). No ethics application was required as they were members of the family of one of the authors. Once we tested the prototype functionality, we directed our efforts to study whether the virtual agent appealed to children, and whether this had repercusions for English vocabulary learning.

*e-mail: e.sandoval@unsw.edu.au
[†]e-mail: anzuetor@ipn.mx
[‡]e-mail: mark.billinghurst@auckland.ac.nz

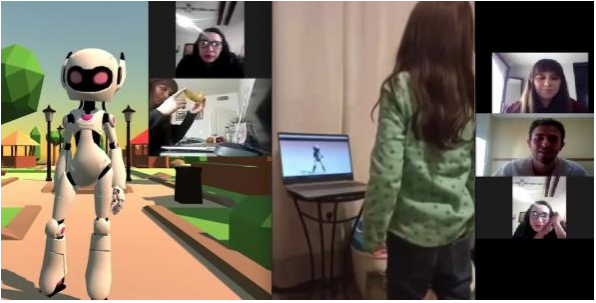

Figure 1: Left: Character implemented in Unity in the second iteration, Right: Child interacting with the IVA assessed by an expert.

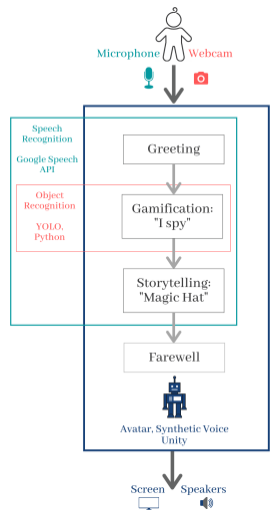

Figure 2: Architecture of the second prototype.

For the second iteration, we recruited six pre-school teachers with a range of work experience and academic qualifications. The second iteration was analyzed using the Heuristic evaluation method proposed by [4]. The evaluation of the second iteration consisted of watching a video of a child interacting with the prototype, and qualitatively evaluating the interaction (see Fig. 1). Experts were asked to answer three questions; (1) What are your views on the prototype? (2) Does the prototype have benefits over conventional methods of learning English? and (3) Do you perceive any possible areas for improvement? These were designed to obtain further qualitative information required for the study.

## 4 RESULTS

In the first iteration, five of the six experts recruited assessed and rated the prototype. The experts regarded it as a novel approach to language learning. They suggested several improvements to make the IVA more engaging to aid the implementation of an interactive teaching strategy. They identified the following usability issues:

a) Unclear and complex dialogues, b) Poor animation quality, c) A Repetitive learning techniques not suitable for children, d) A low level of user engagement and interaction, e) A visual interface leading to a lack of motivation, f) Ineffective feedback which could be a source of frustration for children, g) The prototype lacked kinesthetic stimulation - a key aspect to children learning, and h) Objects had no relation to each other, which hampers learning.

We addressed the usability issues in the second iteration of the IVA prototype. As per Nielsen and Landauer's findings [4], these issues form about 75% of prototype's usability issues. Another test was run to determine whether the issues identified in the first iteration were addressed in the right manner. For instance, we simplified the language used by the IVA and added a more complex story-telling. The same testing and validation protocol was followed to gauge the usability and performance of the second iteration of the IVA. The results obtained suggest that seven out of eight usability issues were solved. The children appeared more engaged during their interaction with the second iteration of the IVA in comparison to the first.

## 5 CONCLUSIONS AND FUTURE WORK.

Our prototype exhibited an acceptable performance in both functionality and usability. It combines artificial intelligence, storytelling and gamification to engage and provide learning for preschool children. Qualitative results demonstrated that usability issues found after the first evaluation were successfully addressed in the second design iteration. End-user testing revealed that both children demonstrated visible excitement while interacting with the second prototype. The experts agreed that using gamification and story-telling for teaching impacts user engagement and the learning process positively. Similarly, these results demonstrate the effectiveness of qualitative testing methodologies in human-computer interaction and human-virtual agent interaction. These outcomes appear to support the notion that the best results come from testing with a small group of expert users as suggested by Nielsen [4].

This project is an early-stage prototype, and further testing and validation is required. We believe that working on technical improvements in reliability and robustness is the first step. Hence, we recommend increasing the types of objects recognized by the artificial vision algorithm and vocabulary used by the speech recognition. Further exploration of state-of-the-art artificial intelligence algorithms might enhance the prototype's performance. The animated agent could use better animations for aesthetic appeal and engagement during interaction. Expansion to mobile devices (tablets, mobile phones) and VR headsets could enhance the prototype portability and accessibility.

Whilst our initial outcomes are promising, further observations of users' perceptions might lead to better design and development approaches in future. The use of typical metrics for the assessment of social agents as Negative Attitudes Toward Robots (NARS), and the Godspeed scale could support quantitatively our current qualitative insights. Finally, we need to establish a better theoretical framework for comparing different language teaching methods and compare with traditional teaching methods.